\documentclass[pre,aps]{revtex4}

\usepackage{rotating}
\begin{document}
\markboth{Dynamics and statistics of heavy particles in
  turbulent flows}{M. Cencini  et al.}

\title{Dynamics and statistics of heavy particles in turbulent flows}

\author{
\centerline{M. CENCINI$^1$, J. BEC$^2$, L. BIFERALE$^3$, G. BOFFETTA$^4$,
    A. CELANI$^5$,}
\centerline{A. S. LANOTTE$^6$, S. MUSACCHIO$^5$ and F. TOSCHI$^7$} 
\vspace{10pt} 
\centerline{$^1$     SMC-INFM c/o Dept. of Physics University of Rome ``La
    Sapienza",}
  \centerline{and CNR-ISC via dei Taurini, 19 I-00185 Roma,
    Italy}
\centerline{$^2$
    CNRS, Lab.\ Cassiop\'{e}e, OCA, B.P.\ 4229, 06304 Nice Cedex
    4, France}\centerline{$^3$ Dept. of Physics and INFN, University
    of Rome ``Tor Vergata'',} 
\centerline{$^4$ Dept. of
    Physics and INFN, University of Torino,}
\centerline{$^5$ CNRS, INLN, 1361
    Route des Lucioles, F-06560 Valbonne, France}
\centerline{$^6$ CNR-ISAC, Sezione di Lecce,
    Str. Prov. Lecce-Monteroni, I-73100 Lecce, Italy}
 \centerline{$^7$ CNR-IAC, Viale
    del Policlinico 137, I-00161 Roma, Italy} \centerline{and INFN,
    Sezione di Ferrara, via G. Saragat 1, I-44100, Ferrara, Italy.} }

\begin{abstract}
  We present the results of Direct Numerical Simulations (DNS) of
  turbulent flows seeded with millions of passive inertial
  particles. The maximum Reynolds number is $Re_{\lambda } \sim 200$.
  We consider particles much heavier than the carrier flow in the
  limit when the Stokes drag force dominates their dynamical
  evolution. We discuss both the {\it transient} and the {\it
    stationary} regimes.  In the transient regime, we study the growth
  of inhomogeneities in the particle spatial distribution driven by
  the preferential concentration out of intense vortex filaments.  In
  the stationary regime, we study the acceleration fluctuations as a
  function of the Stokes number in the range $St \in [0.16:3.3]$. We
  also compare our results with those of pure fluid tracers ($St=0$)
  and we find a critical behavior of inertia for small Stokes
  values. Starting from the pure monodisperse statistics we also
  characterize polydisperse suspensions with a given mean Stokes,
  $\overline{St}$.
\end{abstract}

\maketitle
%%%%%%%%%%%%%%%%%%%%%%%%%%%%%%%%%%%%%%%%%%%%%%%%%%%%%%%
\section{Introduction}
%%%%%%%%%%%%%%%%%%%%%%%%%%%%%%%%%%%%%%%%%%%%%%%%%%%%%%%
Suspensions of dust, impurities, droplets, bubbles, and other
finite-size particles advected by incompressible turbulent flows are
commonly encountered in many natural phenomena and industrial
processes.  These finite-size particles, whose density may differ from
that of the underlying fluid, cannot be modeled as point-like tracers
because of their inertia. Often, they are characterized by the
presence of strong inhomogeneities in their spatial
distribution. Indeed, light (heavy) particles tend to concentrate into
specific regions of the flow characterized by high (low) values of the
vorticity. Such a phenomenon is dubbed `preferential concentration'
\cite{3}. The inhomogeneities then appearing in the particle spatial
distribution are important because they affect the probability to find
close particle pairs and thus influence their possibility to collide,
or to have biological and chemical interactions. Examples showing the
importance of this phenomenon are the formation of rain drops by
coalescence in warm clouds \cite{5,7,8,collins0}, or the coexistence
between several species of plankton in the hydrosphere
\cite{9,10}. Engineering applications encompass optimization of spray
combustion in diesel engines \cite{4} and in rocket propellers
\cite{prop}. Inertial particles are also important for the problem of
dispersion of dust, chemicals or aerosols in the
atmosphere~\cite{11,12}.

Recently, much effort have been devoted to the study of phenomena
related to preferential concentration of inertial particles in
turbulent flows by means of both theoretical
\cite{Falkovich-clustering,Falkovich-Pumir,elperin,simo} and computational
\cite{SE91,Squires,Collins,Wang,Collins2,Collins1} approaches.
Progresses in the characterization of the statistical features of
particle clusters have been achieved by studying inertial particles
evolving in laminar stochastic flows
\cite{stuart,Falkovich-clustering,Bec-Gaw-Horvai,Bec,bccm05} and two
dimensional turbulent flows \cite{Boffetta}.  Experimental results are
reviewed in Ref.~\cite{3}. Recently, experiments~\cite{wara} have been
also realized borrowing techniques from Lagrangian
turbulence~\cite{boden,boden1,ott_mann,Pinton,PintonNJP}.

Considerable less attention has been paid to other aspects of inertial
particle dynamics, and in particular to the statistical properties of
their acceleration in turbulent flows. This is an important issue
relevant, for example, to the development of stochastic models of
particle motion \cite{pdfmod}. It is also interesting to contrast the
statistics of the acceleration of inertial particles with that of
fluid tracers which has been extensively studied by means of
experiments \cite{boden,boden1,sawford,ott_mann,Pinton}, numerical
simulations \cite{bife,biferale2,yeung1,yeung2,IK02,GF01} and
theoretical approaches \cite{hill,GK03,biferale2}. This is even more
interesting on the light of recent experiments \cite{wara} designed to
study inertial particles in turbulent flows by means of Lagrangian
measurements. Concerning fluid tracers, it has been found that the
acceleration probability density function (pdf) displays 
the typical fat tails of highly intermittent signals
\cite{boden,PintonNJP}. The tails of the pdf have been associated to
trapping events into vortex filaments \cite{boden,bife}. Moreover, a
model based on the multifractal description of turbulence has been
proved able to describe the pdf of the acceleration tracers
\cite{biferale2}. Similar questions are still open for inertial
particles. Because of the centrifugal force acting on such particles,
inertia is expected to considerably change the statistics of trapping
events into vortex filaments. It is important to quantify this effect
by looking at the modification of the pdf tails as a function of the
degree of inertia.  Moreover, it is not known how to generalize the
multifractal approach used for passive tracers to the case of passive
particles with inertia.

In this paper we focus on the statistical properties of particles
much heavier than the carrier fluid. We are interested in
investigating, by means of high-resolution Direct Numerical
Simulations (DNS), the combined effects of preferential concentration
and of inertia in determining the statistics of acceleration. With
this aim, we review and extend the results presented in
Ref.~\cite{bbbcclmt05}. In particular, we perform a systematic study
as a function of the Stokes number. Here Stokes is defined as $St =
\tau_s/\tau_{\eta}$, being $\tau_s$ the typical response time of the
particle and $\tau_{\eta}$ the Kolmogorov time of the flow.\\ Most
results concern monodisperse suspensions, i.e. set of particles with a
unique Stokes number. Since laboratory and natural suspensions are
typically polydisperse, we shall extend our analysis to this case
also.\\ Understanding the relevant time scales to reach a stationary
regime is important both for experimental studies -- often limited in
running time -- and for a dynamical description of preferential
concentration.  We present an analysis of the {\it transient regime},
i.e. the time window necessary for the particles to reach a
statistically steady regime, starting from their initial
configuration.

The paper is organized as follows.  In Sec.~\ref{sec:hdyn}, we recall
the equations of motion for the case of inertial particles much
heavier than the surrounding fluid.  A detailed description of the
numerical set up is given in Sec.~\ref{sec:numerics}. In
Sec.~\ref{sec:stat} the validity of the model equations for our
numerical experiments is considered. Moreover, we discuss the
statistical features of preferential concentration both during the
{\it transient} and in the statistically stationary regime. The most
relevant features of particle acceleration statistics for monodisperse
suspensions are discussed in Sec.~\ref{sec:accel}. Section~\ref{sec:poly}
is devoted to extend the previous analysis to polydisperse
cases. Conclusions and perspectives are drawn in Sec.~\ref{sec:con}.

%%%%%%%%%%%%%%%%%%%%%%%%%%%%%%%%%%%%%%%%%
\section{Heavy particle dynamics}
\label{sec:hdyn}
%%%%%%%%%%%%%%%%%%%%%%%%%%%%%%%%%%%%%%%%%
The equations of motion of a small, rigid, spherical particle immersed
in an incompressible flow have been consistently derived from first
principles in Refs. \cite{maxey,maxey2,gatignol} under a certain
number of assumptions. The main working hypothesis is to consider low
particle concentration, so that it is possible to neglect collisions,
particle-to-particle hydrodynamic interactions and the feedback of
particles on the carrier fluid. Another hypothesis is that the
particle radius $r$ is much smaller than any active scale of the
turbulent flow, namely that $r \ll \eta$, where $\eta$ denotes the
Kolmogorov dissipative scale. Finally, the particle Reynolds number $
Re_p = r |V-u|/\nu$, where $|V-u|$ is the typical relative velocity of
the particle with the flow and $\nu$ is the fluid kinematic viscosity,
has to be small, i.e.\ $Re_p\ll1$. These hypotheses lead to describe
the fluid flow surrounding the particle as a Stokes flow and to write
a close differential equation for the motion of the particle
\cite{foot}. The dynamics of a single particle then depends on only
two dimension-less parameters. The first is $\beta = 3\rho_f / (\rho_f
+ 2 \rho_p)$, accounting for the mass density ratio between the
particle, $\rho_p$, and the fluid, $\rho_f$. The second control
parameter is the Stokes number $St = \tau_s/\tau_{\eta}$, where the
particle response time is $\tau_s = r^2/(3\beta\nu)$.  Further,
neglecting gravity, and assuming that the particle is much heavier
than the fluid, the particle velocity $\bm V$ is to leading order a
solution to~\cite{maxey2,Falkovich-clustering}:
\begin{eqnarray}
{d {\bm V} \over dt} &=& 
\beta {D_t u(\bm X,t)}-\frac{{\bm V}(t)-{\bm u}({\bm X}(t),t)}{\tau_s} \;. 
\label{eq:0}
\end{eqnarray}
where $\bm X$ denotes the particle position, ${\bm u}({\bm x},t)$ the
fluid velocity field, and $D_t u(\bm X,t)=\partial_t \bm u+ \bm
u\cdot\bm \nabla \bm u$ is the fluid acceleration at the particle
position. As long as ${D_t u(\bm X,t)}$ does not become too large (see
Sect. \ref{subsec:validity}), the first term in the right-hand side of
(\ref{eq:0}) can be neglected when $\beta \ll 1$, so that the
equations can be further simplified to
\begin{eqnarray}
{d {\bm X} \over dt} &=& {\bm V}(t) \; , \nonumber\\ {d {\bm V} \over
dt} &=& \displaystyle -\frac{1}{\tau_s} \left({\bm V}(t)-{\bm u}({\bm
X}(t),t)\right) \;.
\label{eq:1}
\end{eqnarray}

\noindent It is interesting to compare the evolution equations for
particles (\ref{eq:1}) with that for the motion of fluid tracers
\begin{equation}
\frac{d\bm x(t)}{dt} = \bm u (\bm x(t),t)\,,
\label{eq:tracer}
\end{equation}
which corresponds to the limit $\tau_s=0$ of (\ref{eq:1}). Tracers may
be seen as $St=0$ particles.

\section{Details on the DNS and the data set}
\label{sec:numerics}
%%%%%%%%%%%%%%%%%%%%%%%%%%%%%%%%%%%%%%%%%
The carrier fluid is evolved according to the incompressible
Navier-Stokes equations
\begin{equation}
  \frac{\partial{\bm u}}{\partial t} + {\bm u} \cdot {\bm \nabla} {\bm
    u} = -\frac{{\bm \nabla} p}{\rho_f} + \nu \Delta {\bm u} + {\bm
    f}\,,
\label{eq:2}
\end{equation}
where $p$ is the pressure field and ${\bm f}$ is the external energy
source, $\langle {\bm f}\cdot{\bm u} \rangle = \epsilon$.  These are
solved on a cubic grid with periodic boundary conditions.  Energy is
injected by keeping constant the spectral content of the two smallest
wavenumber shells~\cite{She}. The viscosity is chosen so to have a
well resolved dissipative range, $\eta\approx \Delta x$ where $\Delta
x$ is the grid spacing. We use a fully dealiased pseudospectral
algorithm with 2$^{nd}$ order Adam-Bashforth time stepping. We
performed three sets of runs at resolution $N^3$ with
$N=128,\,256,\,512$, the corresponding Reynolds numbers are
$R_{\lambda} = 65,105,185$ (see also Ref.~\cite{bbbcclmt05}).

Particle dynamics, Eqn. (\ref{eq:1}), is integrated in parallel with
that of the fluid by means of a 2$^{nd}$ order Adam-Bashforth
time-stepping, in which the fluid velocity at particle position is
estimated by linear interpolation.

For each set of runs, the following procedure has been adopted.  We
integrate the Navier-Stokes equations until the flow reaches a
statistically steady state. Once a turbulent configuration for the
flow is obtained, millions of particles and tracers are seeded in the
flow. Their initial positions are uniformly distributed in the volume,
and velocities are chosen equal to the local fluid velocity. Equations
(\ref{eq:1}) for particles and (\ref{eq:tracer}) for tracers are then
advanced in parallel with those for the fluid (\ref{eq:2}). We
consider particles with $15$ different Stokes numbers ranging from
$0.16$ to $3.5$ (see caption of Table~\ref{table:1} for the exact
values), and a set of tracers, $St=0$.  For each set of particles with
a given Stokes number, we save the position and the velocity of $N_t$
particles every $dt = \tau_{\eta}/10$, with a maximum number of
recorded trajectories of $N_t = 5 \cdot 10^{5}$ for the highest
resolution. Along these trajectories we also store the velocity of
the carrier fluid.  At a lower frequency, $\sim 10 \tau_\eta$, we
save the position and velocity of a larger number $N_p$ of particles
(up to $7.5\cdot 10^6$ per $St$, at the highest resolution) together
with the Eulerian velocity field.

A summary of the physical parameters for the run at the highest
resolution, the one discussed in this paper, is given in
Table~\ref{table:1}.
\begin{table*}[ht]
\begin{center}
\begin{tabular}{|ccccccccc|}
  \hline 
  & $R_\lambda$ & $u_{\rm rms}$ & $\varepsilon$ & $\nu$ & $\eta$ & $L$ & $T_E$
  & $\tau_\eta$ \\
  \hline 
     & 185 & 1.4 & 0.94 & 0.00205 &0.010 & $\pi$ & 2.2 & 0.047 \\
   \hline 
\end{tabular}
\end{center}
\begin{center}
\begin{tabular}{|cccccccc|}
  \hline
 & $T_{tot}$ & $T_{tr}$ & $\Delta x$ & $N^3$ & $N_t$ & $N_p$ &
  $N_{tot}$\\ 
\hline   & 12 & 4 & 0.012 & $512^3$ & $5\cdot
  10^{5}$ & $7.5\cdot 10^6$ &$12\cdot 10^7$\\ 
 \hline
\end{tabular}
\end{center}
\vskip 1pt
\caption{DNS parameters. Microscale Reynolds number $R_\lambda$,
  root-mean-square velocity $u_{\rm rms}$, energy dissipation
  $\varepsilon$, viscosity $\nu$, Kolmogorov length scale
  $\eta=(\nu^3/\varepsilon)^{1/4}$, integral scale $L$, large-scale
  eddy Eulerian turnover time $T_E = L/u_{\rm rms}$, Kolmogorov
  timescale $\tau_\eta=(\nu/\varepsilon)^{1/2}$, total integration
  time $T_{tot}$, duration of the transient regime $T_{tr}$, grid
  spacing $\Delta x$, resolution $N^3$, number of trajectories of
  inertial particles for each Stokes $N_t$ saved at frequency
  $\tau_{\eta}/10$, number of particles $N_p$ per Stokes stored at
  frequency $10 \tau_\eta$, total number of advected particles
  $N_{tot}$. Errors on all statistically fluctuating quantities are of
  the order of $10 \%$.  The Stokes number are $St= (0; 0.16; 0.27;
  0.37; 0.48; 0.59; 0.69; 0.80; 0.91; 1.01; 1.12; 1.34; 1.60; 2.03;
  2.67; 3.31)$ }
\label{table:1}
\end{table*}
%%%%%%%%%%%%%%%%%%%%%%%%%%%%%%%%%%%%%%%%%%%%%%%%%%%%%%%%%%%%
\section{Statistical analysis of the data set}
%%%%%%%%%%%%%%%%%%%%%%%%%%%%%%%%%%%%%%%%%%%%%%%%%%%%%%%%%%%%
\label{sec:stat}
First, we check if the approximate equations (\ref{eq:1}) are
consistent with the assumptions used to derive them, in the range of
parameters of our DNS. Second, we study how particles, injected
homogeneously in the domain, reach a statistically inhomogeneous
steady state spatial distribution. Further, we characterize
preferential concentration of particles as a function of the Stokes
number in the stationary regime.
%%%%%%%%%%%%%%%%%%%%%%%%%%%%%%%%%%%%%%%%%%%%%%%%%%%%%%%%%%%%%%%%%%%
\subsection{Limit of validity for the particle equations of motion}
\label{subsec:validity}
%%%%%%%%%%%%%%%%%%%%%%%%%%%%%%%%%%%%%%%%%%%%%%%%%%%%%%%%%%%%%%%%%%%
To validate our working hypotheses, we check that they are
self-consistent with neglecting the term proportional to $\beta$ in
(\ref{eq:0}). In particular, we want to see what are the constraints
on the values of $\beta$ and on the particle radius $r$ that should be
satisfied in order to fulfill the assumption $\beta|D_t\bm u|\ll |\bm
u-\bm V|/ \tau_s$.\\ 
%-----------------
\begin{figure}[th] 
\centerline{\includegraphics[draft=false,scale=0.70]{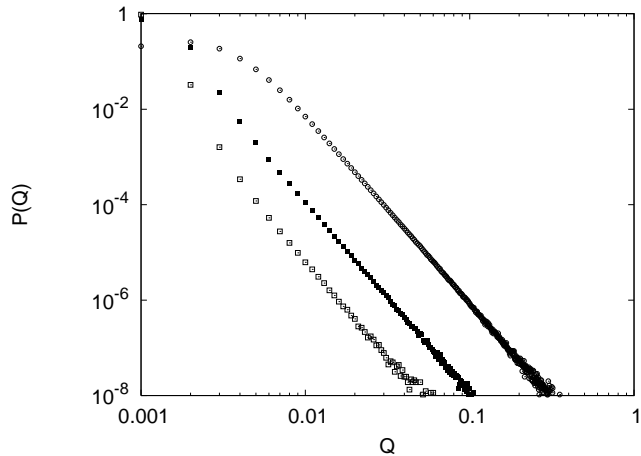}}
\caption{
Log-log plot of the pdf $P(Q)$ vs $Q=\beta \tau_s |d_t u|/|u-V|$, with
$\beta=0.001$ and for three different Stokes numbers $St=0.16, 0.59,
3.31$ from the inner to the outer curve, respectively.}
\label{fig:1}
\end{figure}
%%%%-----------------------------
The first observation is that from the definition of $St$ and from the
dimensional estimation of the Kolmogorov scale
$\eta=(\nu^3/\epsilon)^{1/4}$, one has $ {r/\eta}=\sqrt{3\beta St}\,,$
meaning that for $St\ll 1$ there are no severe constraints on the
value of $\beta$. Differently, for large $St$, in order to have $r\ll
\eta$, one may need to use unphysical small values of $\beta$.  To see
whether this is the case, let us consider a typical example of
parameter values: the case of water in air $\beta\approx 0.001$, and
$St=3$ (which is the largest value used in our simulations). In the
worst case, we have $r/\eta \leq 0.3$ which is still small enough to
justify to neglect the Basset history and the added mass terms, and
the Fax\'en corrections \cite{maxey2}, leading to Eq.~(\ref{eq:0}) for
the particle dynamics. We now show that it is self-consistent to
further neglect the term $\beta D_t\bm u$.  Since in the DNS we do not
have stored the fluid acceleration at particle positions for the whole
trajectory, we assume $D_t \bm u \approx d_t \bm u$, i.e. we estimate
the fluid acceleration at the particle position with the time
derivative of the fluid velocity along the particle path. The validity
of this approximation relies on the fact that the pdf of $d_t \bm u$
that, for large $St$, are very close to those of $D_t\bm u$ (not
shown), meaning that strong acceleration events are well represented.

Figure~\ref{fig:1} shows, for three representative Stokes
numbers, the distribution of the ratio
\begin{equation}
  \label{eq:q}
  Q=\beta \frac{ |d \bm u/dt|} {|( {\bm u}- {\bm
      V})|/\tau_s}.
\end{equation}
As one can see, the statistical weight of events $Q > 0.1$ is very
low, meaning that with a very high probability the equations
(\ref{eq:1}) are a valid approximation. Of course, for density ratios
smaller than $\beta=0.001$ the approximation will work even better. If
one considers for instance sand dust transported in the atmosphere at
an altitude of roughly 5km, $\beta$ is reduced by a factor $1/10$ and
in this case our simulations do not catch any event where $Q > 0.1$.

We thus conclude that in the range of $\beta$ considered above, the
fluid inertia term is also negligible and equations (\ref{eq:1})
correctly describe the motion of small heavy particles in a turbulent
flow at moderate Reynolds number.

%%%%%%%%%%%%%%%%%%%%%%%%%%%%%%%%%%%%%%
\subsection{Transient regimes and formation of preferential concentrations}
%%%%%%%%%%%%%%%%%%%%%%%%%%%%%%%%%%%%%%
It is well known that inertial particles tend to concentrate into
specific regions of the flow.  By performing a local analysis of the
equations of motion it is possible to show that, for small Stokes
numbers, particles heavier than the fluid tend to concentrate onto
strain dominated regions of the flow and escape from rotational
regions, while the opposite is observed for lighter particles
\cite{SE91,Bec,elperin}. Hence, particles do not sample uniformly
the full velocity field and are sensitive to its geometry.

\begin{figure}
\begin{center}
\includegraphics[draft=false,scale=0.52]{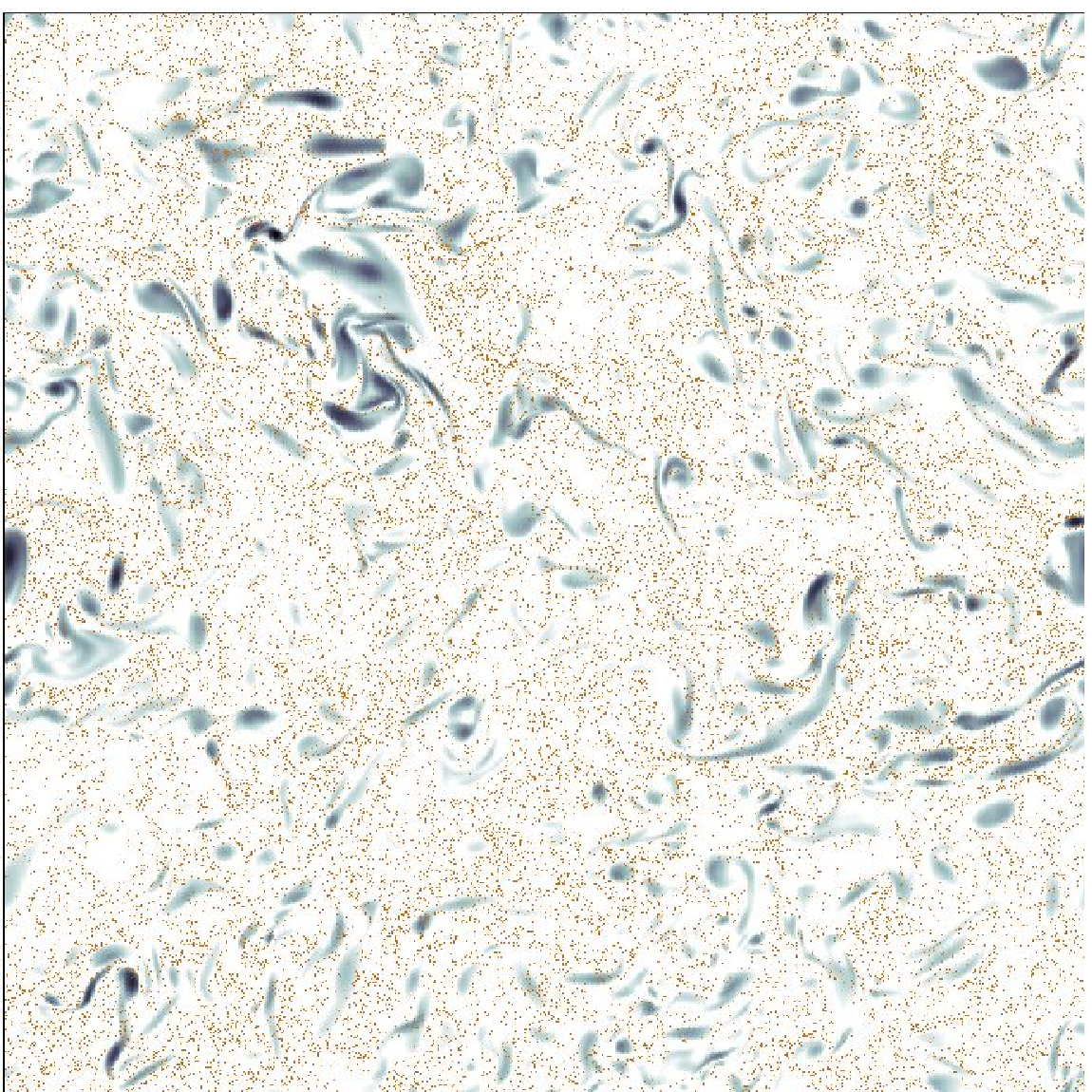} \hfill
\includegraphics[draft=false,scale=0.52]{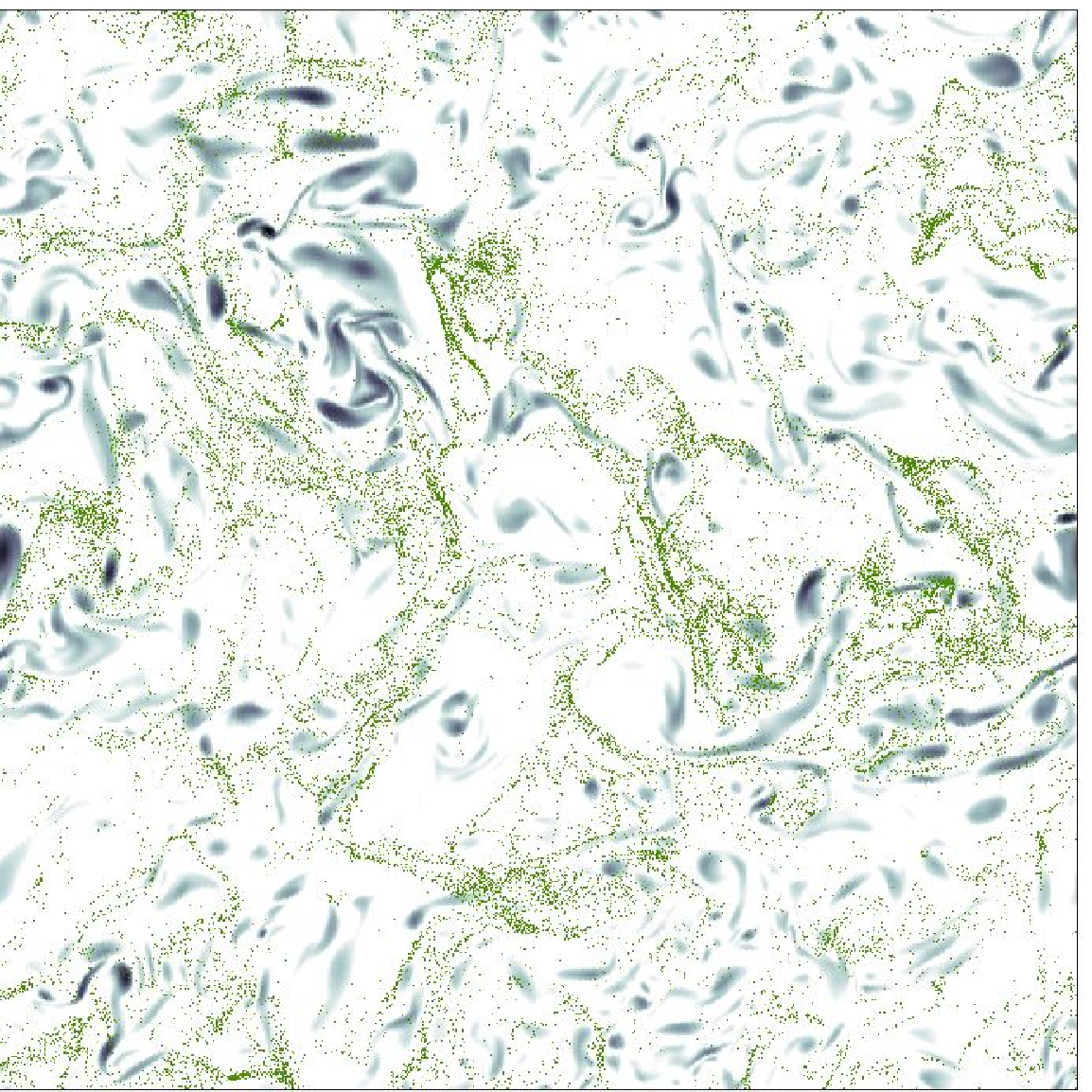}
\end{center}
\caption{(Left) For $St=0.16$, particles positions on a two
  dimensional slice of the simulation box and, in grey scale, the
  amplitude of the vorticity field. (Right) The same for $St=2.03$.}
\label{fig:2}
\end{figure}

Being interested in those properties of particle dynamics that are
closely linked to the flow geometry at the particle locations (such as
the distributions of velocity and of acceleration), one should clearly
understand the bias induced by the correlation between particle
positions and flow structures. Indeed, highly intermittent spots of
the turbulent flow have a strong signature on the acceleration
distribution of simple tracers through the phenomenon of vor\-tex
trapping \cite{boden,bife}. Moreover, at variance with tracers, whose
dynamics is completely determined by that of the velocity field,
inertial particles have their own dynamics. Therefore, for arbitrary
initial data, a relaxation time is needed for them to reach a
statistically stationary motion. Our simulations are started with a
spatially uniform distribution of particles with a vanishing
acceleration (velocity equal to that of the fluid). At sufficiently
large times, the spatial distribution of particles becomes strongly
inhomogeneous and strongly correlates with the flow structure (see
Fig.~\ref{fig:2} for two typical examples of the particle spatial
distribution).
%----------------------------------------
%------------------------------------------------------
As one can see, even for small values of the Stokes number, particles
get less concentrated in those regions where the flow develops strong
vortical structures, represented as dark grey spots on the figure.  To
have insights into the dynamics of heavy particles and to control the
statistical analysis it is important to understand the time scales
involved in the dynamical process of formation of such
inhomogeneities.

A naive inspection of the equation (\ref{eq:1}) for the motion of a
single particle suggests that after few $\tau_s$, the particle
velocity should relax to that of the fluid.  However this relaxation
time is clearly not enough to stabilize the statistical properties of
the particle distribution, and in particular to form statistically
stationary clusters of particles. Indeed, at least for large $St$,
particles organizes in structures with length scales comparable with
the largest ones present in the system (as shown in the right panel of
Fig.~\ref{fig:2}).

\begin{figure}[htbp] 
\centering \includegraphics[draft=false,scale=0.60]{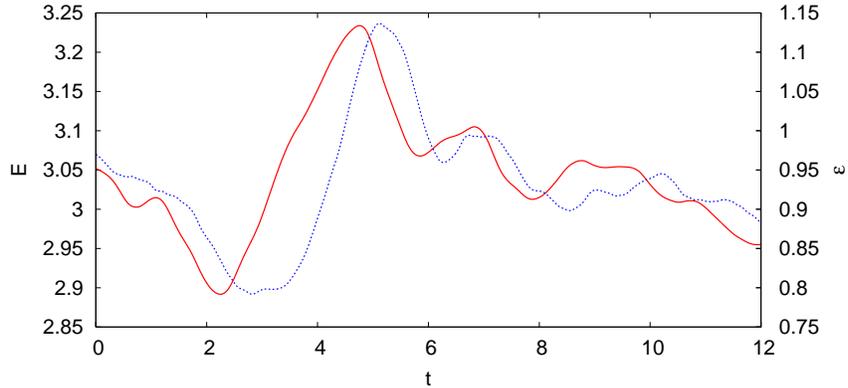}
\caption{Time evolution of the total energy $E=\frac{1}{2}\int d\bm r
  |\bm u(\bm r,t)|^2$ (continuous red line) and of the total energy
  dissipation, $\epsilon$ (dotted blue line).}
   \label{fig:4}
\end{figure}

To quantify the formation of inhomogeneities in the particle
distribution we used two observables. The first one gives a rather
global insight and is based on the coarse-grained mass distribution of
particles. The second observable is local and is related to the
particle distribution at scales of the order of the most violent
events of the turbulent flow, namely the fraction of particles in
rotation and stretching regions of the flow.\\
%------------------------------------------------------------------
\begin{figure}[bh] 
\centering
\includegraphics[draft=false,scale=0.50]{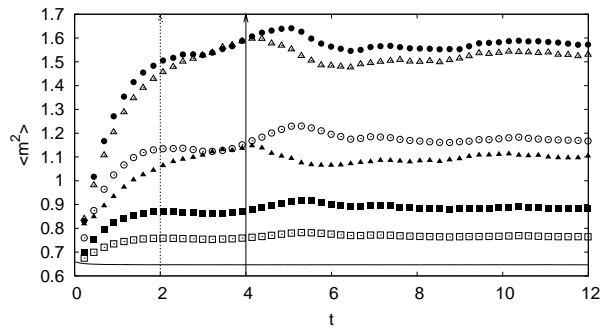} 
\caption{ $\langle m^2 \rangle$ vs time, for particles with
Stokes numbers $St=0$ (bottom solid line), $St=0.16$ (empty boxes), $St=0.27$
(filled boxes), $St= 0.48$ (empty circles), $St=0.90$ (filled circles),
$St=1.60$ (empty triangles) and $St=3.31$ (filled triangles).  The two
vertical arrows indicate the large scale eddy turnover time $T_E
\approx 2.$ (dashed arrow) and the time chosen to be the end of the
transient dynamics $T_{tr} \approx 4$ (solid arrow).
}
   \label{fig:3}
\end{figure}
%--------------------------------------------------------------------
We start the characterization of the inhomogeneous distributions of
particles by measuring the degree of concentration as obtained by
coarse-graining the volume in small cubes of side $\Lambda$, and then
looking at the distribution of the number $m_j$ (or mass) of particles
in the $j$-th cell. Figure \ref{fig:3} represents the behavior of the
second-order spatial moment $\langle m^2 \rangle =
(\Lambda/2\pi)^3\sum_j m_j^2$ of the mass distribution measured on
cubes of side $\Lambda=2\Delta x$. For tracers, which remain
uniformly distributed, $\langle m^2\rangle$ stays constant to the
initial value, as given by a Poisson distribution.  Differently, as
soon as $St > 0$, $\langle m^2 \rangle$ starts to increase in time
until it stabilizes, fluctuating around a mean value. As shown in the
figure all curves reach this steady state only for $t\geq 2\approx
T_E$, i.e.\ for times of the order of the large scale eddy turnover
time.  In performing statistical analysis, for the sake of safety, we
consider that the transient ends at $T_{tr}\approx 2 T_E$ (second
arrow in the figure). Let us notice however that, even in the
statistically steady regime, $ t > T_{tr}$, fluctuations of $\langle
m^2\rangle$ can be rather large and correlated to those of the total
energy (see Fig.~\ref{fig:4}), making the definition of the transition
between unsteady and steady regimes somehow difficult. At lower
resolutions, we obtained quantitatively similar behaviors.  We also
checked that the time scale for equilibration does not depend sensibly
on $\Lambda$ (this remains true only if $\Lambda$ is not too large).\\
%---------------------------------------------------------
%---------------------------------------------------------
We now turn to the second observable based on the particle position
conditioned on the flow local geometry. There are several possible
ways to identify strain or rotation dominated structures in three
dimensions (see Ref.~\cite{haller} for a review)\,: here we
characterize the geometry of the fluid velocity field $\bm u$ by
looking at the eigenvalues of the strain matrix
$\hat{\sigma}_{ij}=\partial_i u_j$. Generally, if the eigenvalues are
all real, the point is said to be hyperbolic, while if they are all
imaginary it is elliptic. In two-dimensions hyperbolic and elliptic
points clearly identify strain and vortical structures,
respectively. In $3d$, pure elliptic points do not exist and the
identification is less straightforward; the hyperbolic or
non-hyperbolic nature of a point can be then identified by looking at
the sign of the discriminant \cite{chong}:
\begin{equation}
\Delta=\left(\frac{{\rm det}[\hat{\sigma}]}{2} \right)^2-\left(
\frac{{\rm Tr}[\hat{\sigma}^2]}{6}\right)^3\,. \label{eq:criteria}
\end{equation}
In deriving (\ref{eq:criteria}), we omitted the term
proportional to ${\rm Tr}[\hat{\sigma}]$ because of incompressibility.
For $\Delta\leq 0$ the strain matrix has $3$ real eigenvalues (strain
 regions), while for $\Delta>0$ it has one real  and $2$
complex conjugate eigenvalues (rotational regions). 
%----------------------------------------------------------
\begin{figure}[t!] 
\centering \includegraphics[draft=false,scale=0.50]{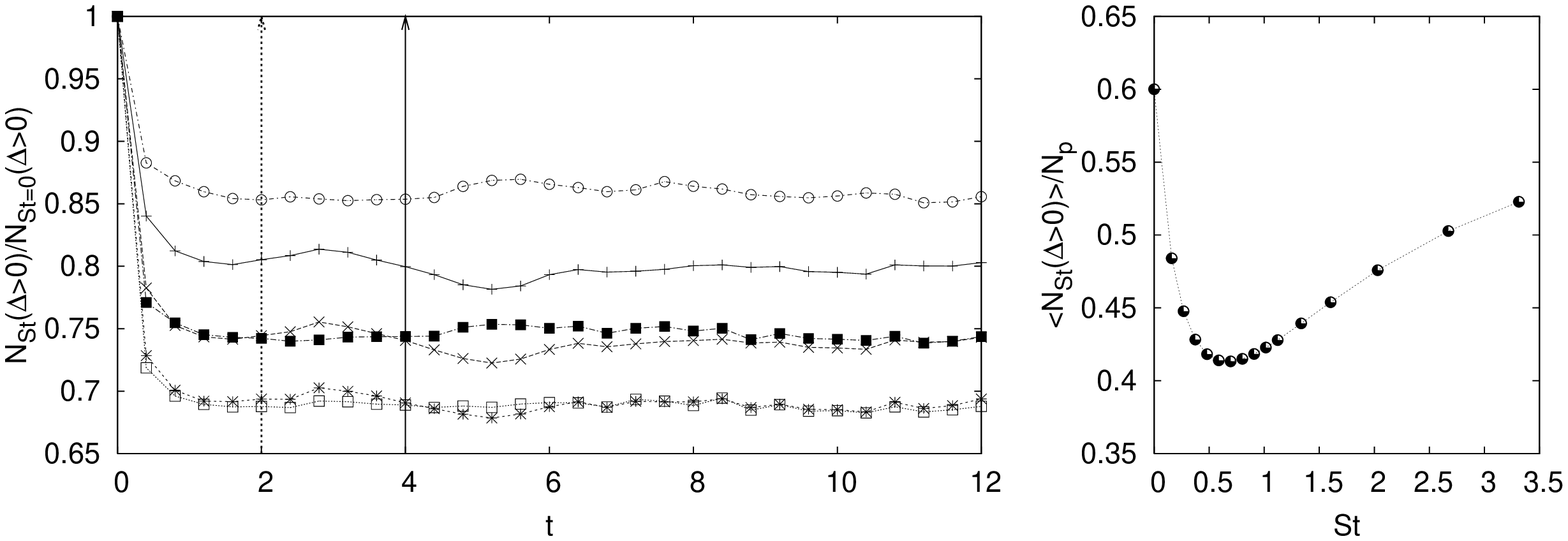}
\caption{Average fraction of particles in the non-hyperbolic regions
  $N_{St}(\Delta> 0)/N_{St=0}(\Delta> 0)$ vs time, for Stokes numbers
  $St=0.16$ (plus), $St=0.27$ (crosses), $St= 0.48$ (stars), $St=0.90$
  (empty boxes), $St=1.60$ (filled boxes) and $St=3.31$ (empty
  circles). The vertical arrows are as in Fig.~\ref{fig:3}. Right:
  average fraction of particles in the non-hyperbolic regions $\langle
  N_{St}(\Delta> 0) \rangle / N_p$ vs $St$. The average has been
  performed considering only steady state data, i.e. for $t\geq
  T_{tr}$.  }
   \label{fig:5}
\end{figure}
%--------------------------------------------
The number of particles $N_{St}(\Delta> 0)$ in $\Delta > 0$ regions is
represented in Fig.~\ref{fig:5} as a function of time and for
different Stokes numbers.  We chose here to normalize this quantity by
the number $N_{St=0}(\Delta> 0)$ of tracers in non-hyperbolic regions
to smooth out the instantaneous fluctuations of hyperbolic and
non-hyperbolic points in the flow. At $t=0$ this ratio is $1$ for all
values of $St$ because particles are uniformly injected in the
domain. As time goes on, it follows a very fast drop from the initial
value. Unfortunately, the strong fluctuations, correlated to those of
the total energy (Fig.~\ref{fig:4}), do not permit to unambiguously
clarify if the time required for $N_{St}(\Delta> 0)/N_{St=0}(\Delta>
0)$ to stabilize is or not of the same order of that needed by
$\left\langle m^2 \right\rangle$. However, according to
Fig.~\ref{fig:5}, it does seem that the correlation with the flow geometry
settles on a time scale shorter  than that needed to stabilize
particles mass moments. It would be interesting to test this point, by
studying the simultaneous evolution of particles uniformly injected,
and particle already stabilized with the surrounding flow.  This would
help to disentangle spurious effects induced by the fluctuation of the
global properties of the flow.

We conclude this section by commenting the right panel of
Fig.~\ref{fig:5}, where we show the average number of particles
normalized by the total number $N_p$, $\langle N_{St}(\Delta> 0)
\rangle / N_p $ in the stationary regime.  As previously observed
\cite{bbbcclmt05}, it appears that preferential concentration is
non-monotonic in $St$, but there is an ``optimal'' Stokes number for
which the effect is maximal, here $St \simeq 0.55$.  This behavior can
be understood as follows. To have particles concentrated into specific
flow structures, their response time must be fast enough to allow them
to follow the flow evolution. Clearly large $St$ particles are unable
to do that and tend to de\-correlate from the flow geometry; though
their large scale distribution may still be strongly inhomogeneous (as
indeed observed, Fig.~\ref{fig:2}). On the other hand, for $St\ll 1$
particles behaviour is closer and closer to that of tracers, whose
positions do not correlate with the flow geometry. As we will see in
next section the balance between these two effects has a strong
signature in the probability distribution of particle acceleration.

%%%%%%%%%%%%%%%%%%%%%%%%%%%%%%%%%%%%%
\section{Statistical properties of the acceleration}
\label{sec:accel}
\begin{figure}
\centering \includegraphics[draft=false,scale=0.70]{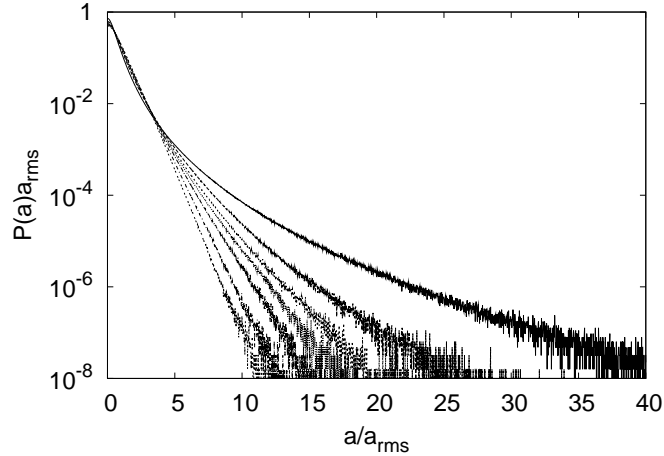}
\caption{Normalized acceleration pdf's for a subset of Stokes values
  ($St=0,0.16,0.37,0.58,1.01,2.03,3.31$ from top to bottom) at
  $R_{\lambda}=185$. The darker line corresponds to tracers, $St=0$.}
\label{fig:6}
\end{figure}

%%%%%%%%%%%%%%%%%%%%%%%%%%%%%%%%%%%%%
One of the most striking features that characterize tracers is the
very intermittent distribution of the acceleration
\cite{boden,boden1,Pinton,bife}\,: fluctuations as strong as $80$
times the root mean square acceleration $a_{rms}$ reflect the tendency
of tracers to be trapped into vortical structures
\cite{boden,bifejot1,bifejot2}. We would like to understand if this
property is shared also by inertial particles and what are the main
features of the acceleration fluctuations, at changing the importance
of inertia.
%Here, we briefly
%review and describe with more details some results presented in
%\cite{bbbcclmt05}. 
A natural way to proceed is to start from the
equations of motion. In the statistically stationary state, the formal
solution of Eq.(\ref{eq:1}) is:
\begin{eqnarray}
{\bm V}(t)&=&\frac{1}{\tau_s}\int_{-\infty}^t e^{-(t-s)/\tau_s} {\bm
  u}({\bm X}(s),s) \, ds \nonumber
\label{eq:3}\\
{\bm a}(t)&=&\frac{1}{\tau_s^2}\int_{-\infty}^t e^{-(t-s)/\tau_s} 
[{\bm u}({\bm X}(t),t)-{\bm u}({\bm X}(s),s)]\, ds \;.
\label{eq:4}
\end{eqnarray}
It is clear that the kernel $e^{-(t-s)/\tau_s}$ acts as a low-pass
filter on the fluid velocity differences, suppressing frequencies
larger than the inverse of the Stokes time $\tau_s^{-1}$. The larger
is the Stokes number $St$, the more important this effect should
be. On the other hand, we have indications that the limit of
vanishing Stokes number is {\it singular}, due to the clusterization
of particles onto fractal sets \cite{Bec,bccm05}. This should have
consequences on the acceleration statistics also. \\ In
Fig.~\ref{fig:6}, the probability density functions for the 
particles acceleration are plotted for different Stokes numbers and
compared with that of tracers. As we could expect, acceleration
fluctuations become less and less intermittent as $St$
increases. To disentangle different effects, it is useful to
analyze the two limiting cases of small and large Stokes numbers,
separately.\\
%%%%%%%%%%%%%%%
%%%%%%%%%%%%%%%%%%%%%%%%%
\noindent {\em At small $St$}, i.e. $\tau_s \ll \tau_\eta$, the fluid
velocity along the trajectory evolves smoothly in time and the
acceleration reduces to ${\bm a(t)} \simeq d_t \bm u({\bm X}(t),t)$,
i.e. to the derivative of fluid velocity along the inertial particle
trajectory. In the limit of vanishing $St$, this is indistinguishable
from the fluid acceleration $D_t {\bm u}({\bm X}(t),t)$ evaluated at
particle positions and one expects that the particle acceleration
essentially coincides with the fluid acceleration. There is however
one major difference with the tracer case\,: inertial particles
preferentially concentrate inside regions with low vorticity (see
right panel of Fig.~\ref{fig:5}).
%%%%%%%%%%%%%%%%%%%%%%%%%%%%%%%%%%%%%%%
\begin{figure}
\centering
\includegraphics[draft=false,scale=0.70]{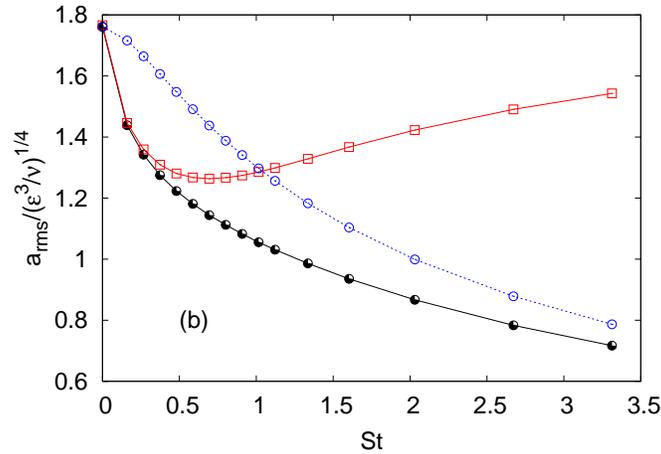}
 \caption{Comparison between  $a_{\rm rms}$
   (fill circles) as a function of Stokes with the acceleration of
   the fluid tracer conditioned to be on the particle position,
   $\langle (D_t{\bm u})^2 \rangle^{1/2}$ (empty boxes). The last curve
   (empty circles), approaching $a_{\rm rms}$ for large $St$, is the
   one obtained from the filtered tracer trajectories, $a^F_{\rm
     rms}$.}
\label{fig:7}
\end{figure}
%%%%%%%%%%%%%%%%%%%%%%%%%%%%%%%%%
In Fig.~\ref{fig:7}, we plot the results for $a_{\rm rms}$, as a
function of $St$. Notice the singular effect of the inhomogeneous
spatial distribution which results in a drastic reduction of $ a_{\rm
  rms}$ already for small $St$. This behavior is even more pronounced
for the fourth moment of the acceleration \cite{bbbcclmt05}. To better
understand the importance of preferential concentration, we have
measured the tracers acceleration $\langle (D_t{\bm u})^2
\rangle^{1/2}$ conditioned on the spatial positions of the inertial
particles: the result is also plotted in Fig.~\ref{fig:7}. The
agreement of the two curves at small $St$ confirms the validity of the
previous argument. \\ However, at increasing $St$, the curves start to
deviate. While $a_{\rm rms}$ monotonically decreases in $St$, the
tracer acceleration conditioned on the particle positions has a
minimum for $St \approx 0.5$ close to the maximum of preferential
concentration, eventually recovering the value of the unconditioned
tracers for larger $St$. For increasing $St$, we indeed know that
inertial particles explore the small scale structures of the flow more
and more homogeneously and a different mechanism is responsible for
the reduction of $a_{\rm rms}$.\\

\noindent {\em At large $St$}, i.e. $\tau_s \gg \tau_\eta$, we indeed expect
the filtering effect to become more and more important.  Again, we
start from tracers: for each solution, $\bm x(t)$ of
Eq. (\ref{eq:tracer}), we define a new velocity, $\bm u^F$, obtained
from filtering the fluid velocity along the particle trajectory over a
time window of the size of the order of the Stokes time:
\begin{equation}
\label{eq:filt}
{\bm u^F}(t)=\frac{1}{\tau_s}\int_{-\infty}^t e^{-(t-s)/\tau_s} {\bm
u}({\bm x}(s),s) \, ds,
\end{equation}
then the filtered acceleration is given by $ {\bm a}^F = d{\bm
  u^F}/dt$. In figure~\ref{fig:7} the acceleration variance of
particles $a_{\rm rms}$ is compared with that obtained from tracers in
the above manner, without any additional spatial conditioning. The
curves corresponding to $a_{\rm rms}$ and to $ a^F_{\rm rms}$ become
closer and closer as $St$ increases, supporting the conjecture that
preferential concentration for $St>1 $ becomes less important. \\ For
intermediate $St$ a non trivial interplay between the two above
mechanisms takes place. A model, even qualitative, able to bridge the
gap between the two limits in still unavailable.\\
%%%%%%%%%%%
\begin{figure}
\centering \includegraphics[draft=false,scale=0.49]{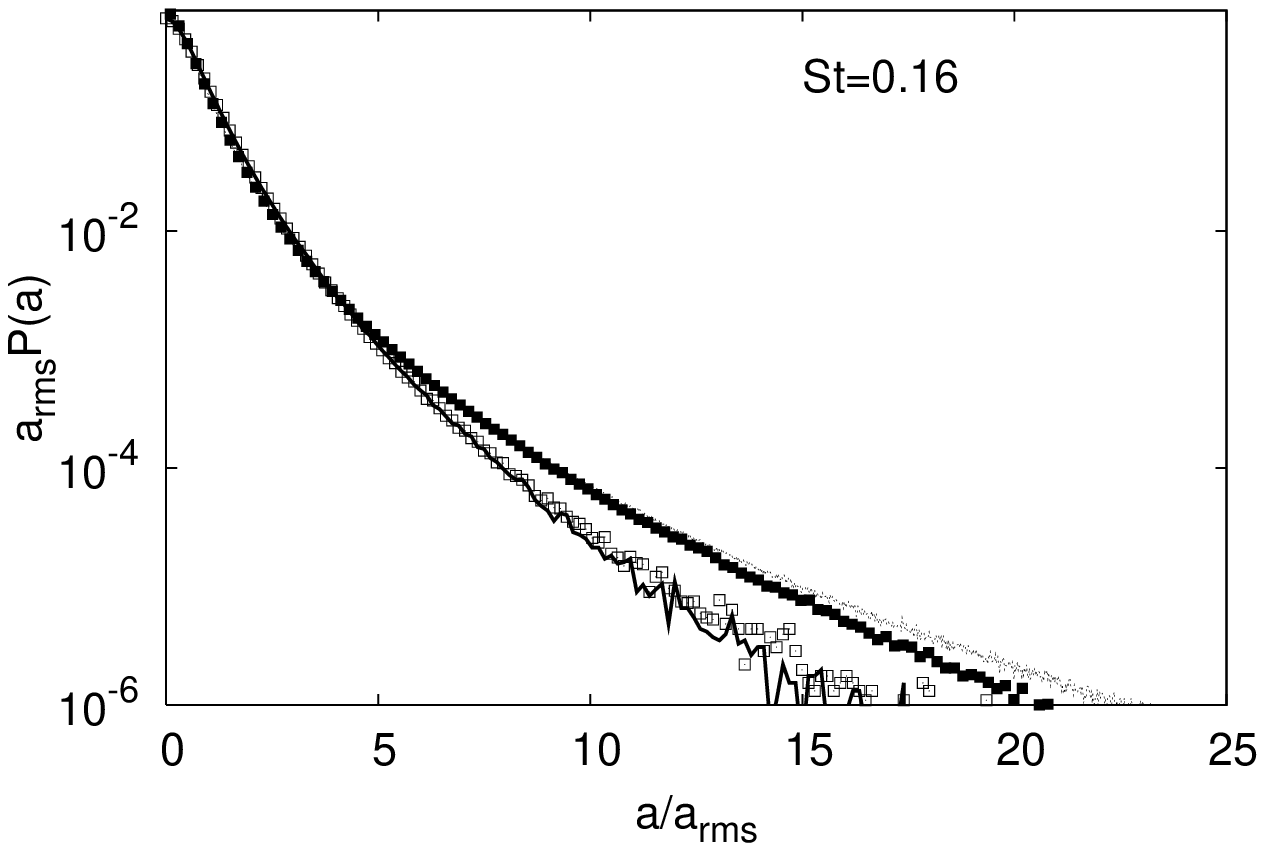}
\includegraphics[draft=false,scale=0.49]{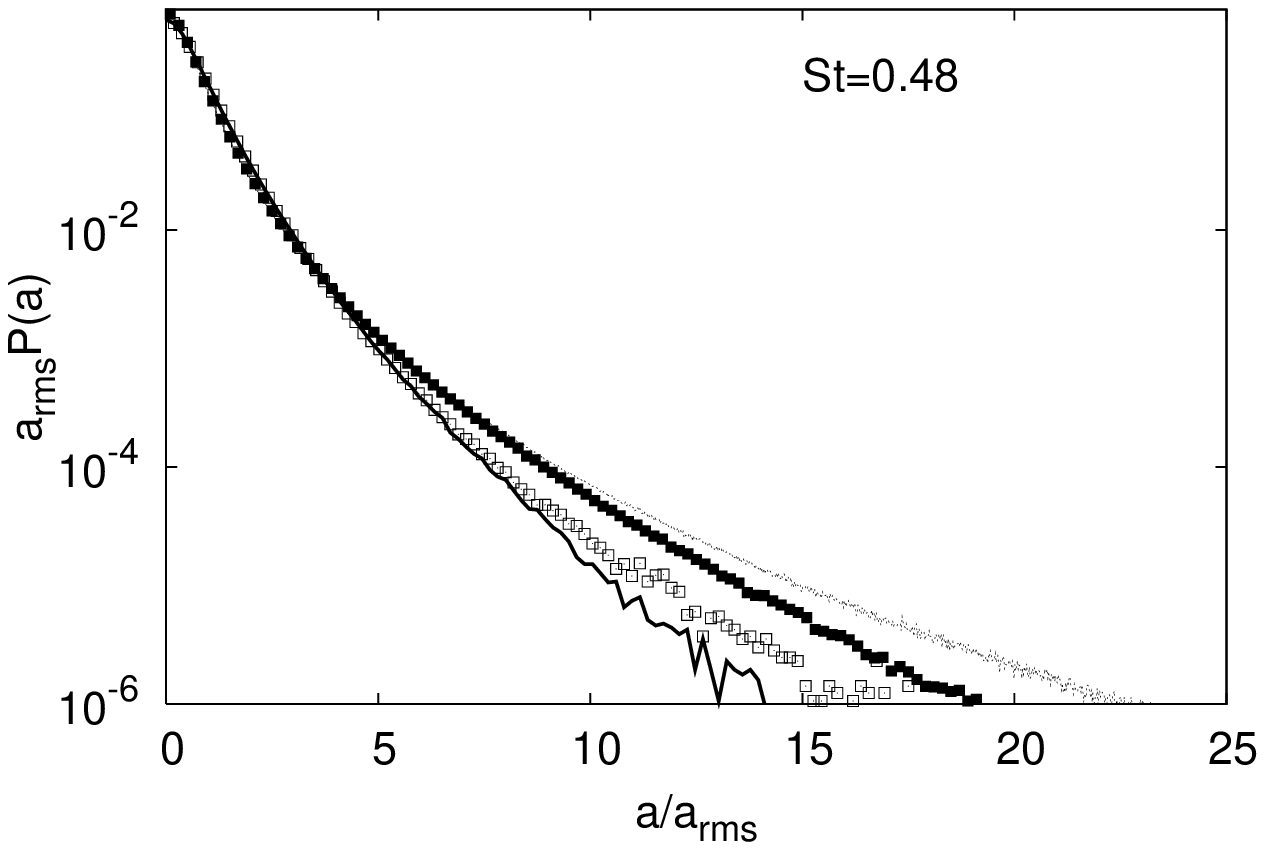}\\
\includegraphics[draft=false,scale=0.49]{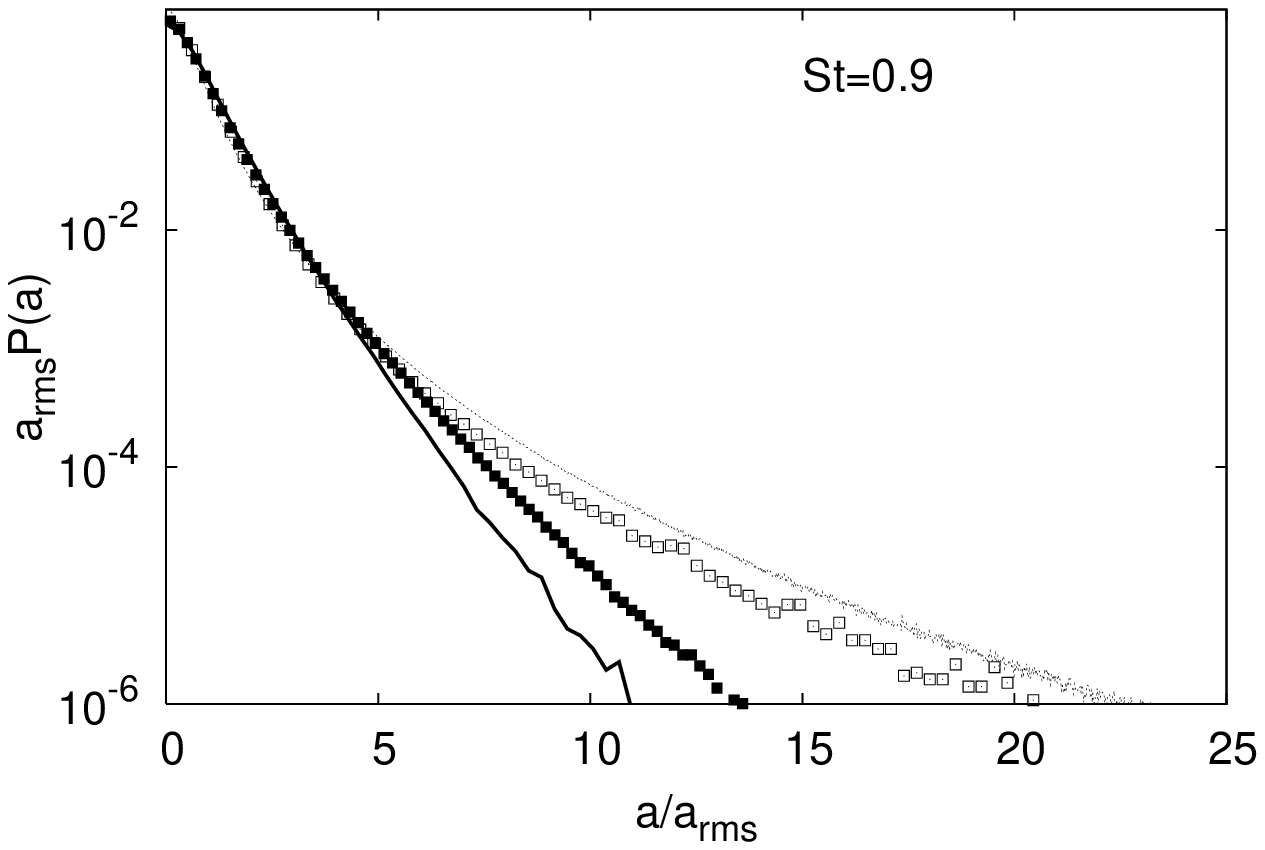}
\includegraphics[draft=false,scale=0.49]{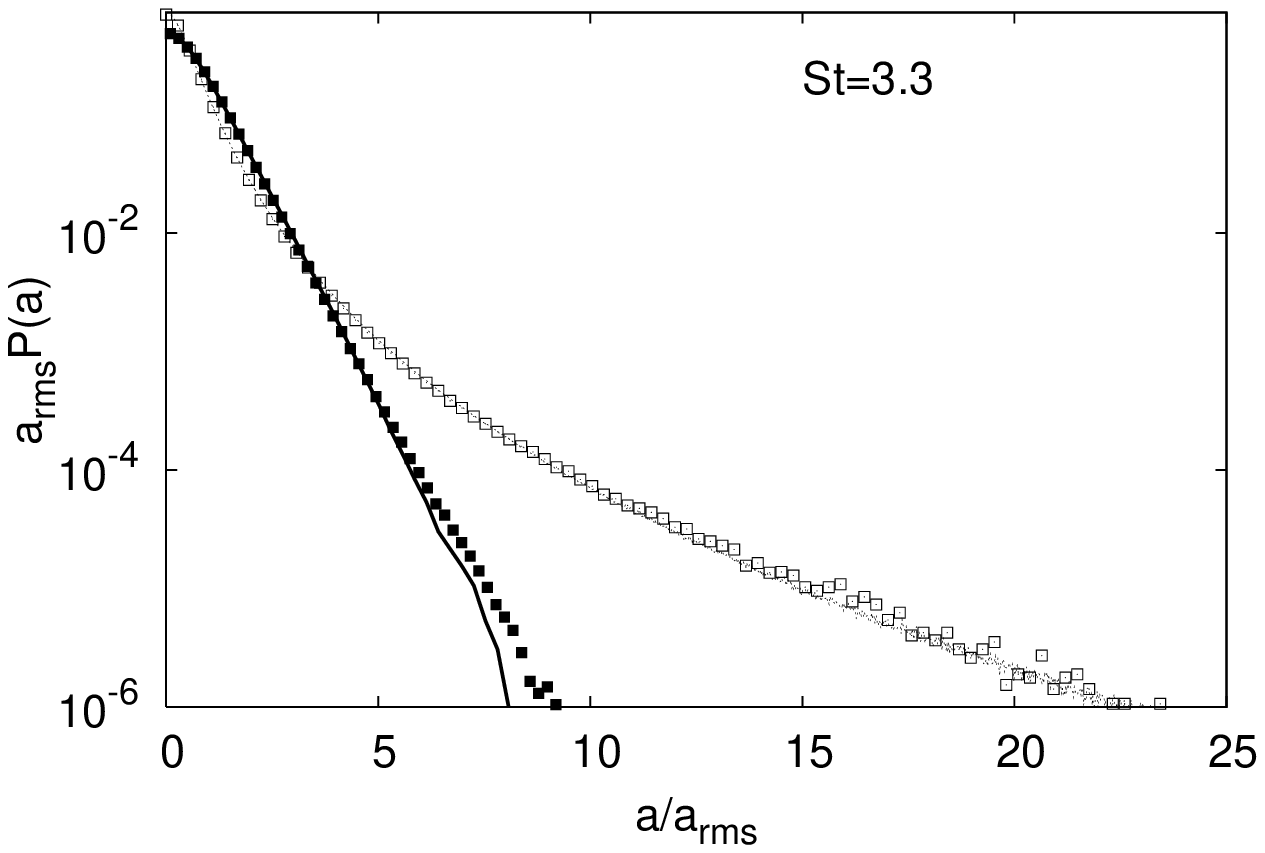}
\caption{ In each panel the following four curves are shown: the
  normalized acceleration pdf for inertial particles (solid line), the
  acceleration pdf of fluid tracers $D_t\bm u$ conditioned to particle
  positions (empty boxes), the acceleration pdf of filtered fluid
  trajectories $ {\bm a}^F$ (filled boxes) and the unconditioned
  acceleration pdf for fluid tracers(dotted line). Note that the
  filtered acceleration $ {\bm a}^F$ pdf, which is for small $St$
  practically coincident with the (unfiltered and unconditioned) fluid
  acceleration pdf, becomes closer and closer to the particle
  acceleration pdf as $St$ increases. The opposite is observed for the
  tracer acceleration conditioned to particle positions.}
\label{fig:8}
\end{figure}
%%%%%%%%%%%%%%%
In the limit of small and large $St$ the qualitative trend of the
pdf's can be captured by the same arguments as those used for $a_{\rm
  rms}$.  In Fig.~\ref{fig:8} we compare the normalized pdf particle
acceleration with those obtained by using the tracer acceleration
measured on the particle position, $D_t\bm u$, and the filtered tracer
trajectories, $a^F$, respectively. For the smallest Stokes number
$St=0.16$, the overlap between the conditional tracer acceleration and
the particle acceleration is almost perfect, while the filtered one is
very close to the unconditioned tracer acceleration pdf. The opposite
is observed for the largest Stokes number, $St=3.31$. Notice that the
(normalized) conditional tracer acceleration pdf is very close to the
unconditioned one, meaning that preferential concentration is no
longer effective on the small scale statistics for large $St$. For two
intermediate $St$ values, we observe the transition from preferential
concentration dominated regime to the filtering dominated one.

We conclude this section by going back to the instantaneous
distribution of particles for small and large $St$ (see
Fig.~\ref{fig:2}). Both pictures present an inhomogeneous spatial
distribution for the inertial particles but for different reasons. At
small $St$ (left panel), inhomogeneity is mostly related to the fact
that particles escape from large vorticity regions. For such small
response times $\tau_s$, particles are able to follow the flow
variations but the very small scales properties of the fluid are
sampled in a biased way. This explains the remarkable agreement between
particle acceleration statistics and fluid tracers conditioned on
particle positions. At larger Stokes number, the inhomogeneity is less
correlated to the fine structures of the flow. Indeed we observe void
regions whose typical size can be much larger than the Kolmogorov
length scale and comparable to the inertial range scales. In spite of
the clear inhomogeneous distribution of particles, the small scales, $
r \ll \eta$, are sampled more homogeneously than at smaller $St$. This
is coherent with the fact that the conditional tracer acceleration
statistics is very close to the unconditioned one.
%%%%%%%%%%%%%%%%%%%%%%%%%%%%%%%%%%%%%%%%%%%%%%%%%%%%%%%%%%%
\section{Acceleration of polydisperse suspensions}
\label{sec:poly}
%%%%%%%%%%%%%%%%%%%%%%%%%%%%%%%%%%%%%%%%%%%%%%%%%%
Suspensions generally involves particles distributed over different
sizes. This is true not only for phenomena observed in nature, as for
aerosol distribution in the atmosphere, but also for laboratory
experiments. In some cases, particles can also considerably change
their size with time, as it happens for water droplets in warm clouds
\cite{pruppa}, so varying their Stokes number. Direct numerical
simulations allow to study inertial particles in the idealized case of
a given Stokes number, as we have done previously, but also permit to
test phenomenological approaches for polydisperse flows. In this
section, we study what happens to the acceleration pdf when particles
have a non trivial size distribution, or equivalently when the Stokes
number of a given suspension is a random variable with a distribution
${\cal Q}(St)$. This is particularly important to have clean and
possibly unambiguous comparison of numerical simulations results and
laboratory experiments.\\
%-------------------------
\begin{figure}
\centering \includegraphics[draft=false,scale=0.7]{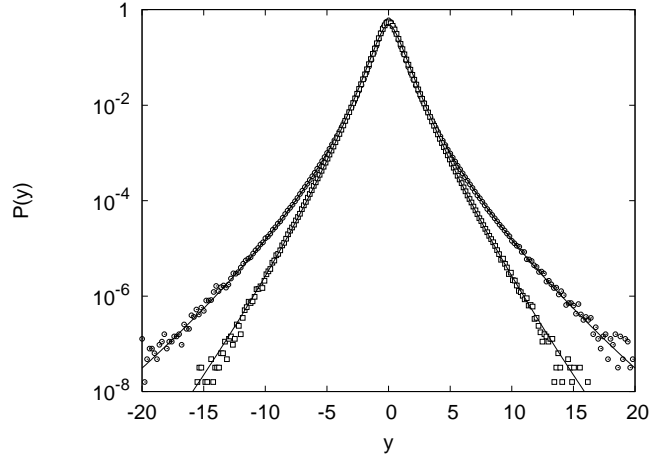}
\caption{Comparison of the acceleration pdf from DNS data (symbols)
  and the fit (solid line) of eq.(\ref{eq:fit}) for two monodisperse
  cases: for $St=0.27$ (outer curves) and $St=1.01$ (inner curves).}
\label{fig:9}
\end{figure}
%----------------------------------
To model the acceleration pdf ${\cal P}_{poly}(a)$ in polydisperse
flows, we first need a functional form for the pdf, ${\cal P}(a|St)$,
of particles with a given Stokes number $St$. Second we make a
convolution of the latter distribution with ${\cal Q}(St)$. Building
up a phenomenological model for the acceleration pdf of monodisperse
suspension valid at different Stokes number is not an easy goal. For
tracers, the multifractal formalism, without free parameter, is able
to predict an acceleration pdf which is in excellent agreement with
DNS data \cite{biferale2}. However any simple extension of such
approach to heavy particles does not seem to work, because of the
almost singular nature of the small $St$ limit and of the complex
interplay between preferential concentration and time filtering. In
the lack of such a model, the first step is done by fitting the
acceleration pdf at changing the Stokes number. At
small values of the acceleration, i.e.~ in the pdf core, we expect to
have a Gaussian-like profile ${\cal P}(a|St)\sim \exp(-a^2)$; while
for large fluctuations $a \gg a_{rms}$, we should recover a
distribution with stretched exponential tails, ${\cal P}(a|St) \propto
\exp(-a^{\gamma})$. A simple functional form for the pdf, ${\cal
  P}(a|St)$ of the normalized acceleration $y=a/{a_{rms}(St)}$
satisfying these two requirements is:
\begin{equation}
\tilde{\cal P}(y|St) \equiv a_{rms}\,{\cal P}(y\,a_{rms}|St)=
\exp{\left\{-\frac{c_1\,|y|^{2}}{1 + c_3\,y^{c_2}} + c_4\right\}}.
\label{eq:fit}
\end{equation}
Here, to have a simple fitting procedure, we have used four free
parameters $c_i(St)$ with $i=[1,\dots,4]$, although two of them may be
fixed by the requirement of having the pdf normalized to unit area and
unit variance.  Of course, the most important parameter is $c_2(St)$
which fixes the Stokes number dependency of the far tail exponent
$\gamma = 2 - c_2$.\\
%---------------------------------------------
\begin{figure}
\centering
\includegraphics[draft=false,scale=0.50]{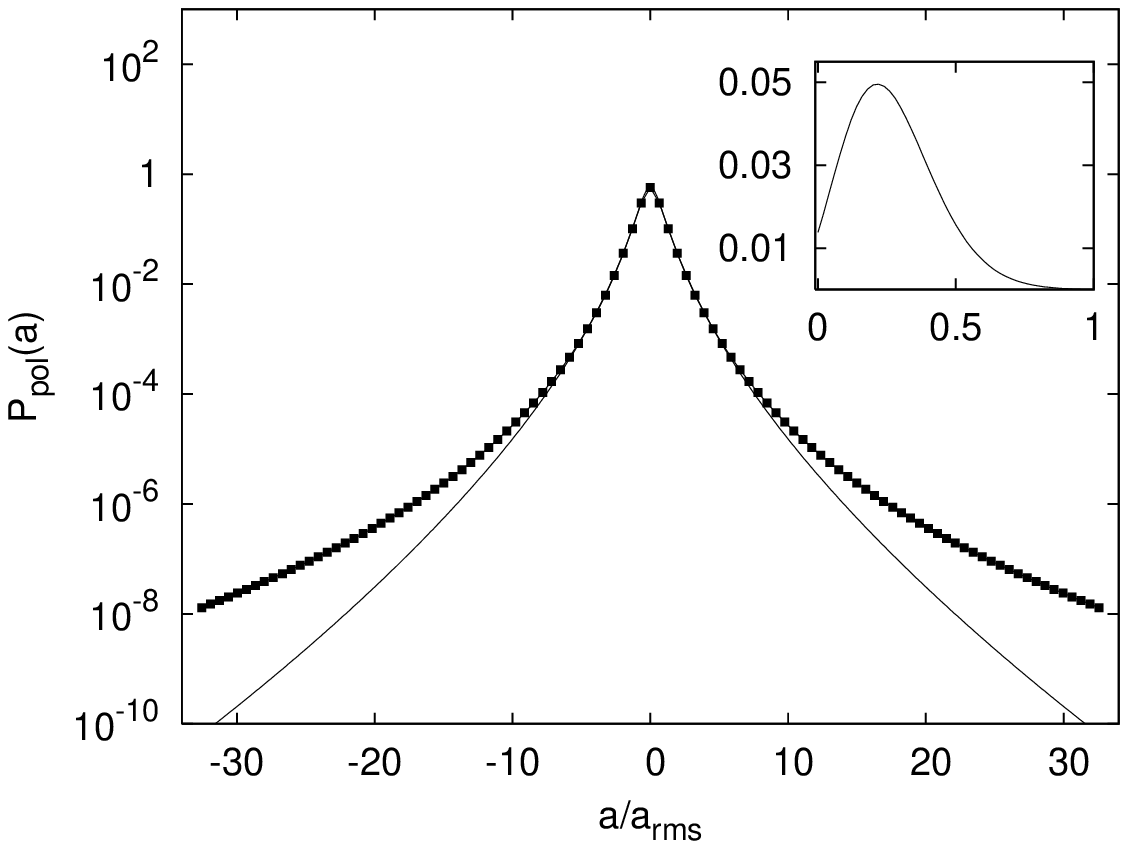}
\includegraphics[draft=false,scale=0.50]{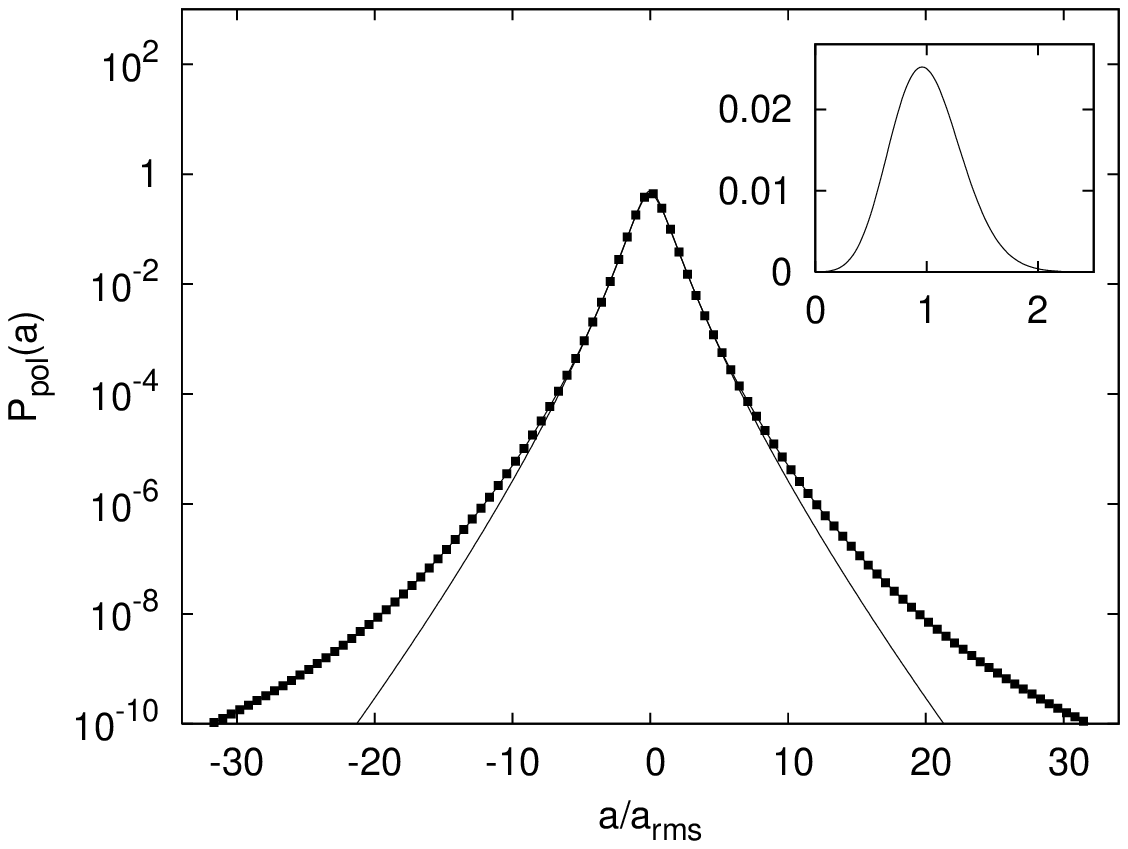}
\caption{Comparison between the acceleration pdf for a polydisperse
  case (filled squares) of mean Stokes value ${\overline{St}}$, and a
  monodisperse case (solid line) at the Stokes number
  $St={\overline{St}}$. In the inset, the Poisson distribution ${\cal
    Q}(St)$ of mean Stokes value ${\overline{St}}$. Left panel refers
  to case with the Stokes number ${\overline{St}}=0.27$, while right
  panel refers to ${\overline{St}}=1.01$.  }
\label{fig:10}
\end{figure}
%-------------------------------------
The acceleration pdf measured in our DNS has been fitted with
expression (\ref{eq:fit}) for all available $St$.  The values for the
four free parameters are those reported in Table~\ref{tab:fit}. They
all gently, but non monotonically, vary with the Stokes number. In
particular, the exponent $c_2$, which controls the fat tails, goes
from about $\sim 1.6$, for the smallest Stokes, to $\sim 1.2$, for the
largest available one. This is in agreement with the tendency towards
a less intermittent distribution for particles with a larger
inertia. From Fig.~\ref{fig:9}, we can see that the fitting formula
(\ref{eq:fit}) reproduces well the acceleration distribution (here
shown for $St=0.16$ and $St=1.01$ only).\\ To assess the variations for
a polydisperse suspension, we make the convolution between the
acceleration pdf conditioned on the Stokes number ${\cal P}(a|St)$ and
the Stokes number distribution ${\cal Q}(St)$\,:
\begin{equation}
\label{eq:poly}
{\cal P}_{pol}(a) = \int dSt\,\, {\cal P}(a|St) {\cal Q}(St).
\end{equation}
A possible choice for ${\cal Q}(St)$ is a Poisson-like distribution\,:
for any available Stokes number $St$, we generate a Poisson pdf with
mean value $\overline{St}=St$ and we insert it as test curve to
evaluate the convolution (\ref{eq:poly}). In Fig. \ref{fig:10} we
compare the polydisperse pdf (\ref{eq:poly}) obtained with a
Poissonian distribution, ${\cal Q}(St)$, with a given $\overline{St}$
and a monodisperse distribution with $ St = \overline{St}$. We do it
for two different average Stokes number, namely $\overline{St} = 0.27,
1.01$. \\ As expected, the effect of a random distribution for the
Stokes number is much more important for the tails than for the
cores. We recall also that, in the case of monodisperse suspensions
the pdf tails are the most sensitive to $St$ variations. Hence, any
polydisperse suspensions characterized by a particle size distribution
skewed towards small values of the Stoeks numeber should develop more
intermittent tails than those expected on the basis of their mean
Stokes value.
\begin{table*}[t]
\centering
\begin{tabular}{ccccccccc}
\hline 
St & $0$ & $0.16$ & $0.27$ & $0.37$ & $0.48$ & $0.59$ & $0.69$ & $0.80$ \\
\hline 
$c_1$ & 2.68 & 2.17 & 2.24 & 2.31 & 1.67 & 1.64 & 1.83 & 1.66 \\
$c_2$ & 1.55 & 1.43 & 1.38 & 1.35 & 1.36 & 1.33 & 1.29 & 1.28 \\
$c_3$ & 0.79 & 0.75 & 0.83 & 0.91 & 0.62 & 0.63 & 0.76 & 0.68 \\
$c_4$ &-0.41 &-0.53 &-0.53 &-0.53 &-0.65 &-0.67 &-0.64 &-0.68\\
\hline
\end{tabular}
\vspace{0.2cm}
\centering
\begin{tabular}{ccccccccc}
St  & $0.91$ & $1.01$ & $1.12$ & $1.34$ & $1.60$ & $2.03$ & $2.67$ & $3.31$\\
\hline 
$c_1$ & 0.83 & 1.64 & 1.15 & 1.08 & 1.18 & 1.95 & 0.89 & 0.76\\
$c_2$ & 1.40 & 1.25 & 1.30 & 1.28 & 1.24 & 1.11 & 1.23 & 1.27\\
$c_3$ & 0.24 & 0.69 & 0.42 & 0.39 & 0.46 & 1.00 & 0.30 & 0.21 \\
$c_4$ &-0.98 &-0.69 &-0.82 &-0.85 &-0.81 &-0.67 &-0.90 &-0.94 \\
\hline 
\end{tabular}
\label{tab:fit}
\centering
\caption{The values of the fitting parameters $c_i$ at changing the
  Stokes number. Notice that only two out of the four parameters are
  indeed independent because the pdf must be normalized to have unit
  area and unit variance. The fitted functional form is not meant to
  have a solid phenomenological base. It is used, as simple as it is,
  to estimate the importance of polydisperse suspensions with respect
  to monodysperse ones.}
\end{table*}
%%%%%%%%%%%%%%%%%%%%%%%%%%%%%%%%%%%%%%%%%%%%%%%%%%%%%%%%%%%%%%%%%%%%%%%%%
\section{Conclusions and perspectives}
%%%%%%%%%%%%%%%%%%%%%%%%%%%%%%%%%%%%%%%%%%%%%%%%%%%%%%%%%%%%%%%%%%%%%%%%%
\label{sec:con}
Direct numerical simulations, even at moderate Reynolds number,
represent a valuable tool for the study of the Lagrangian motion of
heavy particles. Here we have described the specific set up of our
numerical study, analyzed the importance of preferential concentration
and particle clustering as a function of the Stokes number. Moreover,
we have explored the transient dynamics which precedes the settling of
the particle motion onto a statistically steady state in terms of the
particle distribution. This may be particularly important for real
observations and laboratory experiments, where long time records of
particle motion are hardly attainable. We have also reviewed and
extended some recent results about the acceleration pdf of inertial
particles, at varying the importance of inertia. The main conclusions
are (i) preferential concentration plays an {\it almost singular} role
at small Stokes, since even a small inertia is sufficient to expel
particles from vortical regions where the strongest acceleration
fluctuations are experienced; (ii) for small Stokes, a good
quantitative agreement between the inertial particle acceleration and
the conditioned fluid tracer acceleration is obtained; (iii) at large
Stokes, the main effects is filtering of the velocity induced by the
response Stokes times. \\The formulation of a phenomenological model
able to describe the inertial particle acceleration as a function of
both Stokes and Reynolds numbers is still beyond our reach. An easy
fitting procedure has however been proposed for polydisperse
suspensions, the most common experimental situation. For polydisperse
cases, we show that the simple approach of characterizing the
suspension in terms of its average Stokes number without taking into
account the Stokes number distribution can give a systematic bias,
particularly if the particle distribution is skewed towards small
Stokes values. It would be extremely important to test the proposed
approach with experimental data. Finally it would be interesting to
numerically investigate $St\ll 1$ particles to  get a better
understanding of the almost singular behavior that has been 
observed for the acceleration. \bigskip

We acknowledge useful discussions with S. Ayyalasomayajula,
E. Bodenschatz, A. Gylfason, G. Falkovich and Z. Warhaft. This work
has been partially supported by the EU under the research training
network HPRN-CT-2002-00300 ``Stirring and Mixing''. Numerical
simulations have been performed thanks to the support of CINECA
(Italy) and IDRIS (France) under the HPC Europa project
(RII3-CT-2003-506079).  We thank also the ``Centro Ricerche e Studi
Enrico Fermi'' and N.~Tantalo for technical support.


\begin{thebibliography}{99}
\bibitem{foot} Generally, these hypothesis are also
  sufficient to neglect the contributions of the Basset history and
  the added mass terms, and the Fax\'en corrections.
\end{thebibliography}
\end{document}